\documentclass{aa}
\usepackage{amsmath,graphicx,amssymb}
\usepackage{natbib}
\bibpunct{(}{)}{;}{a}{}{,}

\newcommand\pdra{\object{$\varphi$\,Dra}}
\newcommand{\zav}[1]{\left(#1\right)}
\newcommand{\hzav}[1]{\left[#1\right]}
\newcommand\intvidpo{\!\!\int\limits_{\begin{array}{c}\text{\scriptsize
visible}\\[-2mm]\text{\scriptsize surface}\end{array}}\!\!}

\newlength\staretab

\makeatletter
\def\sgn{\mathop{\operator@font sgn}\nolimits}
\makeatother

\begin{document}

\title{Modelling of variability of the CP star $\varphi$ Draconis}

\author{
    M.~Prv\' ak\inst{1} \and
    J. Li\v ska\inst{1} \and
    J.~Krti\v{c}ka\inst{1} \and
    Z.~Mikul\'a\v sek\inst{1} \and
    T.~L\"uftinger\inst{2}
 }

\offprints{M.~Prv\' ak,\\  \email{prvak@physics.muni.cz}}

\institute{
    Department of Theoretical Physics and Astrophysics,
    Masaryk University, Kotl\'a\v rsk\' a 2, CZ-611\,37 Brno, Czech~Republic
    \and
    Institute of Astrophysics, University of Vienna,
    T\"urkenschanzstra\ss e 17, 1180 Vienna, Austria
}

\date{Received}

\abstract
{ 
    The presence of heavier chemical elements  in stellar atmospheres influences
    the spectral energy distribution (SED)  of stars.  An uneven surface
    distribution  of  these elements,  together  with  flux  redistribution  and
    stellar  rotation, are  commonly  believed  to  be  the primary causes of the
    variability of chemically peculiar (CP) stars.
}
{ 
    We aim  to model  the  photometric  variability of the CP star $\varphi$ Dra
    based on  the assumption of  inhomogeneous  surface  distribution of heavier
    elements and compare  it to the  observed  variability  of the star. We also
    intend to identify  the   processes   that  contribute   most  significantly
    to its photometric variability.
}
{ 
    We use a grid of TLUSTY model  atmospheres and the SYNSPEC code to model the
    radiative  flux  emerging  from  the  individual surface elements of \pdra{}
    with  different chemical compositions.  We integrate the emerging  flux over
    the visible surface of the star at different phases  throughout  the  entire
    rotational period  to synthesise  theoretical  light curves  of the  star in
    several spectral bands.
}
{ 
    The synthetic light curves  in the visible and in the near-UV regions are in
      very  good  agreement   with  the  observed   variability  of  the  star.
    The lack of usable  far-UV  measurements  of the star precludes  making  any
    conclusions about the  correctness of our model in this spectral region.  We
    also  obtained  194  new  \textit{BVRI}  observations  of  $\varphi$ Dra and
    improved its rotational period to $P=1\fd716500(2)$.
}
{ 
    We show that the inhomogeneous distribution of elements, flux redistribution,
    and rotation  of the star are fully capable of explaining the stellar
    variability in the  visible and the near-UV regions. The flux redistribution
    is  mainly  caused  by  bound--free  transitions of silicon and bound--bound
    transitions of iron.
}


\titlerunning{Modelling of $\varphi$\,Dra variability}
\authorrunning{M.~Prv\' ak et al.}
\maketitle

\section{Introduction}
    The  star  \pdra~(HD~170000 = HR~6920) is one  of  the  brightest  known  CP
    stars.  It  was  mentioned  in  the  work  of \citet{maury1897}, where it is
    classified as  {\sc Group VII} and  marked  peculiar.  It is
    a  multiple hierarchical system~\citep[e.g.][]{tokovinin2008} 
    containing  visual (A, B) and  spectroscopic (Aa, Ab) binary pairs.
    
    It has been known for many years that the  brightest  component  Aa  is
    variable  in  some  spectral  regions \citep[e.g.][]{kukarkin1969, tencolour}.
    The  observed   variability  is   strongest  in  the  far-ultraviolet  (UV)
    region  of  the  spectrum.   In  the  near-UV  and  in  the  visible  the
    variability  is  much  weaker,   and  it  is  in  antiphase  to  the  far-UV
    variations  of  the  star  \citep{jamar1977}.    This  kind  of  variability
    has   been   observed  in   some  other  CP  stars  (e.g.  $\alpha^2$~CVn),
    and  it  is  typically  explained by the inhomogeneous  horizontal   surface
    distribution  of  heavier  elements,   spectral  energy  redistribution,  and
    rotation of the star \citep[e.g.][]{molnar1973}.

    Bound--bound transitions  in  the  absorption  lines of heavier elements, as
    well as bound--free transitions in the areas with  increased  metallicity
    cause absorption of a significant portion  of the  radiative  energy  in the
    far-UV part of  the spectrum.  This  inevitably leads to an  increase in  the
    temperature  in the stellar atmosphere  (the  backwarming  effect),  and the
    radiative flux in the near-UV  and in the visible increases, so that the
    overall  energy  emitted  by  the  star  remains  unchanged  and the energy
    equilibrium is maintained.  As the star rotates,  we can observe variability
    in  individual  spectral  regions  as  the  areas  with  different  chemical
    composition move across the visible stellar surface.

    It  is  possible to derive  surface  abundance maps with observations of the
    line profile variations of various chemical elements throughout  the  entire
    rotational period  of  the star. To this end, techniques like  Doppler 
    imaging (DI) and magnetic Doppler  imaging (MDI) are  commonly  used with
    success \citep[e.g.][]{rice,khokhlova,piskunov,kochukhov,luftinger,
    luftinger2,bohlender}.

    Availability  of  sufficiently  detailed  atomic data and high quality model
    atmospheres  \citep[e.g.][]{tlusty}  allows  us  to  accurately compute the
    emergent flux from stellar atmospheres  with various  chemical compositions.
    Putting these together with the abundance maps,  it is  possible  to compute
    the  emergent  flux  from  individual  regions  of the atmospheres of the CP
    stars  and,  integrating  the  flux  over  the  visible  surface of the star
    repeatedly  for   different   rotational   phases,   model  the  photometric
    variability in various spectral regions. This approach may serve  as  a test
    of the  theory,  our comprehension of the physical processes taking place in
    stellar atmospheres, the accuracy of atomic data and model  atmospheres, and
    the correctness of the abundance maps.

    This  method  has  been  used  in  the  past  to  show  that the photometric
    variations of several CP stars are caused by inhomogeneous  distribution  of
    various chemical elements. For example,  the light variations of HD~37776 in
    the \textit{u,\,v,\,b}, and $y$ bands of the Str\"omgren photometric  system
    are  caused  by spots of helium and silicon \citep{krticka2007}.  Similarly,
    \citet{krticka2009} showed that  most of the variability  of HR~7224  can be
    explained by inhomogeneous distribution of silicon and iron. In the case  of
    $\varepsilon$~UMa,  iron and chromium are responsible  for  the  variability
    \citep{shulyak2010}. \citet{krticka2012} explained   most of the variability
    of CU~Vir by the inhomogeneous distribution of silicon and iron, but  they
    were unable to  fully  explain  the  variations  of  the  star in the region
    of 2000--2500~\AA.

    We  intend  to  use  the  same  technique  to model  the light variations of
    \pdra\ using the abundance maps obtained from \cite{kuschnigthesis}. Our
    preliminary results were published in \citet{prvak2014}.

\section{Modelling of the SED variability}

    Our aim is to compare synthetic light curves, derived  from our models based
    on maps  constructed from the  spectroscopy  obtained  at  Observatoire  de
    Haute-Provence  using the {\sc  Aur\'elie}  spectrograph
    \citet[see also \citealt{kuskon}]{kuschnigthesis} to  the   ten-colour
    photometry   by
    \citet{musielok1980}. Because of the large time gap  between these  two data
    sets, we had to  improve  the  rotational  period  of  the star and test its
    stability.

    \subsection{Our photometric observations}
        The  studied  star  \pdra\  with  its  $V=4.22$\,mag  belongs  to the
        brightest  chemically  peculiar  stars in the northern hemisphere,
        which negatively influenced its photometric monitoring.  Despite this,
        we  have  collected  673  individual  photometric  measurements covering
        forty years (1974--2013) -- see Table~\ref{oc} for the list of
        observations. The last item represents new measurements, which  we obtained 
        using  a small refractor.
        The observations were performed at the Masaryk University Observatory (2
        nights) and at the Brno Observatory and Planetarium (15 nights) from May
        to  July 2013.  Only  the  12  best  nights  were  used  for  the period
        analysis. Observing equipment consisted of a  photographic  lens Jupiter
        3.5/135\,mm (lens speed/focal length) and  a   Moravian  Instruments CCD
        camera  G2-0402 with Johnson-Cousins \textit{BVRI} photometric filters.
        The  angular  resolution  was  only  14$''$/pixel  and  thus it  was not
        enough  for angular separation of the fainter visual B component (actual
        distance around 0.5$''$, see Fig.~\ref{fig-system}). The CCD images
        were calibrated  in  a standard
        way (dark frame  and  flat  field  corrections). The \textsc{C-Munipack}
        software\footnote{http://c-munipack.sourceforge.net/}(Motl 2009)   which
        is based on DAOPHOT \citep{stetson1987}  was used for this processing and  for  differential  ensemble  photometry.  Stars  HD~169666 and
        HD~171044 were used as comparison stars.

    \subsection{Observed light variation: Rotational period}

        \begin{figure}[t]
            \centering
            \resizebox{0.9\hsize}{!}{\includegraphics{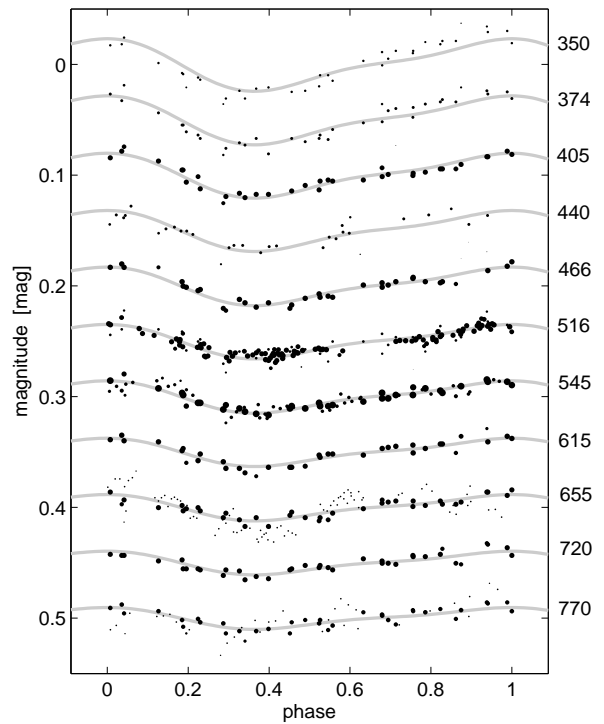}}
            \caption{
                Light curves of \pdra\ in 11 photometric colours
                with  effective  wavelengths  from 350  to  770 nm (on the
                right) plotted versus the  new  ephemeris with zero phase at the
                maximum  (Eq.~\eqref{efem}).   Grey  full  lines  are  the  fits
                calculated  by  the  simple  model  of  \pdra\ light variability  
                (Eq.~\eqref{LCmodel}). Light curves are  vertically  shifted  to
                better  display  the  light  variability.  Areas  of  individual
                symbols are  proportional  to the weight of the measurement.}
            \label{krivky}
        \end{figure}
        \begin{table}
             \centering
                \caption{
                List  of  \pdra\ photometry  used  for  the improvement  of  the
                ephemeris   (see Eq.~\eqref{efem}).}
                \label{oc}
                \begin{tabular}{ccccl}
                    \hline
                    Years  & Type & $N$ & \textit{O-C}$_{\rm{new}}$ & Source  \\
                    \hline
                    1974--5 & 10 col. &375 & 0\fd00095(55) & \citet{musielok1980} \\
                    1991--2 & $H_{\mathrm{p}}$ & 104 & $-$0\fd005(9) & \citet{esa97} \\
                    2013    & \textit{BVRI} & 194 & 0\fd006(14) & this paper \\
                    \hline
                \end{tabular}
                \tablefoot{$N$ denotes the number of measurements and
                \textit{O-C}$_{\rm{new}}$ are \textit{O-C} values derived using
                the   ephemeris based on Eq.~\eqref{efem}.}
        \end{table}
    
        We modelled the observed photometric variations in the range from 350 to
        770 nm by  a  simple  phenomenological  relation (see Fig.~\ref{krivky})
        assuming a linear ephemeris,
        \begin{eqnarray}
            & \displaystyle m_{j}(t_i,\lambda_j)=\,\displaystyle\overline{m}_j
                +\hzav{A_1+ A_2\zav{\frac{545\,\mathrm{nm}}{\lambda_j}-1}}\,F(\vartheta_i),
                \label{LCmodel} \\
            & F(\vartheta,a_1,a_2)=\sqrt{1-a_1^2-a_2^2}\,\cos(2\pi\vartheta)+a_1
                \cos(4\pi\vartheta)                                   \\
            & \displaystyle +\frac{2\,a_2}{\sqrt{5}}\hzav{\sin(2\pi\vartheta)
                -\frac{1}{2}\sin(4\pi\vartheta)},\quad \vartheta_i=
                \frac{t_i-M_0}{P}, \nonumber  
        \end{eqnarray}
        where $m_j(\vartheta_i,\lambda_j)$  is a magnitude of the $j$-th
        data  subset  obtained  in  the  photometric  filter  with the effective
        wavelength  $\lambda_j$ in  nm,  $t_i$  is  JD$_{\mathrm{hel}}$
        time  of the $i$-th measurement,  $\overline{m}_j$ is the mean magnitude
        of the  $j$-th data subset,  $A_1$ and $A_2$ are amplitudes expressed in
        magnitudes, $\vartheta$ is a phase function, $F(\vartheta,a_1,a_2)$ is a
        normalised second order harmonic polynomial with a  maximum at the phase
        $\varphi = \mathrm{floor}(\vartheta) = 0$,   $a_1$  and  $a_2$  are 
        dimensionless parameters,  $M_0$  is  the  time  of the basic extremum
        of the function $F(\vartheta)$, and $P$ is the period. 
    
        The parameters were found by non-linear  robust regression
        \citep[a brief  description  is  given in][]{RR}, which effectively
        eliminates outliers.
        After some iteration we found the following linear ephemeris,
        \begin{equation}
            \mathrm{JD}_{\mathrm{hel\,max}}(E)=2\,445\,175.023(9)+1\fd716500(2)
                \times E, \label{efem}
        \end{equation}
        where $E$ is an integer. The other model parameters are $a_1=0.214(3)$,
        $a_2 = -0.477(3)$,   $A_1 = -0.0134(3)$\,mag, and   $A_2 = -0.0149(3)$\,mag.
        The total amplitude  at  345\,nm  is  0.047\,mag; the amplitude at 
        770\,nm is 0.020\,mag.
 
       \begin{figure}[t]
           \centering
           \resizebox{1\hsize}{!}{\includegraphics{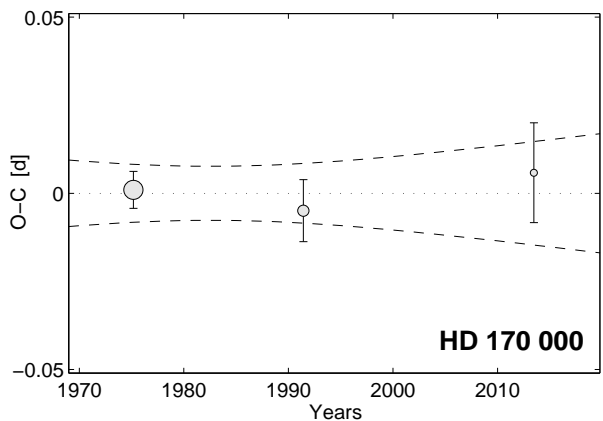}}
           \caption{
           \textit{O-C}  diagram  for \pdra\ calculated versus a
               new linear ephemeris 
               (Eq.~\eqref{efem}).
               Areas  of the  circles are
               proportional  to  the  weight of individual \textit{O-C}  values.
               The  dashed lines denote  1-$\sigma$ uncertainty of a light curve
               maximum time prediction.}
            \label{OC}
        \end{figure}

        We recalculated  the rotational phases  of the abundance maps  based  on
        these  new values.  The improved  ephemeris  has resulted in a  shift of
        the  modelled  light  curves  by   about  0.16  in phase compared to the
        ephemeris used by \cite{kuschnigthesis}.  There  are  no signs of period
        variations (see Fig.~\ref{OC}).

    \subsection{Model atmospheres and synthetic spectra}

        For  the  modelling  of  the variability   of  \pdra,   we   used    the
        stellar  parameters   adopted  from  \cite{kuschnigthesis}.  A  list  of
        these  parameters,  including the  ranges of abundances of the elements,
        for   which   the   abundance   maps   were   available,  is   given  in
        Table~\ref{tab-param}.   Here,   the    abundances    are    given    as
        $\varepsilon_\text{el}=\log(N_\text{el}/N_\text{tot})$.
        
        \begin{table}[t]
            \caption{Parameters of \pdra\ from spectroscopy
                     \citep{kuschnigthesis}.}
            \label{tab-param}
            \begin{center}
                \begin{tabular}{lc}
                    \hline
                    Effective temperature ${T_\mathrm{eff}}$ & ${12\,500}$\,K \\
                    Surface gravity ${\log g}$ (cgs)         & ${4.0}$        \\
                    Rotational inclination ${i}$             & ${60^\circ}$   \\
                    Rotational velocity projection
                    $v_\text{rot} \sin i$ & $95\,\text{km}\,\text{s}^{-1}$    \\
                    Helium abundance      & $-2.7<\varepsilon_\text{He}<-2.1$ \\
                    Silicon abundance     & $-4.4<\varepsilon_\text{Si}<-3.1$ \\
                    Titanium abundance    & $-7.1<\varepsilon_\text{Ti}<-6.5$ \\
                    Chromium abundance    & $-6.5<\varepsilon_\text{Cr}<-5.2$ \\
                    Iron abundance        &$-4.9<\varepsilon_\text{Fe}<-3.6$  \\
                    \hline
                \end{tabular}
            \end{center}
        \end{table}

        \begin{table}[t]
            \caption{
            Abundances used for model atmospheres.}
            \label{tab-abun}
            \begin{center}
                \begin{tabular}{lllll}
                    \hline
                    Element & \multicolumn{4}{l}{Abundances}                \\
                    \hline
                    He      &  $-2.56$  &           &           &           \\
                    Si      &  $-4.60$  &  $-4.10$  &  $-3.60$  &  $-3.10$  \\
                    Ti      &  $-6.85$  &           &           &           \\
                    Cr      &  $-6.70$  &  $-6.20$  &  $-5.70$  &  $-5.20$  \\
                    Fe      &  $-5.10$  &  $-4.60$  &  $-4.10$  &  $-3.60$  \\
                    \hline
                \end{tabular}
            \end{center}
        \end{table}

        We  calculated  a  grid  of  model  atmospheres  for  several  different
        chemical compositions  in order to cover  the range of abundances across
        the surface of the star (the individual values of  the  abundances  used
        are listed  in  Table~\ref{tab-abun}).  The  atomic data were taken from
        \citet{tlusty},  and   are   based  on  \citet{mendo},  \citet{maslo93},
        \citet{kur22},     \citet{nah96,nah97},        \citet{bau96,bau97},
        and Kurucz (2009)\footnote{http://kurucz.harvard.edu}.

        Given the very weak variability  of helium  and titanium abundances and
        their low values (see Table \ref{tab-param}), we decided to exclude
        these
        two elements from the adopted abundance grid and use only a single value
        (the mean value) for each of them  throughout  the computations.    This
        allowed us  to significantly reduce the number of model atmospheres   to
        be calculated.

        The  model   atmospheres   were   computed   using   the   TLUSTY   code
        \citep{tlusty}. They were all  LTE plane-parallel  models,  because  the
        NLTE effects do not have significant influence  on the light variability
        of   CP   stars   \citep{krticka2012}.  We  also  assumed  the effective
        temperature  ${{T}_\mathrm{eff}}$  and  the  surface gravity  ${\log g}$
        to   be   constant   across    the    entire   stellar   surface    (see
        Table~\ref{tab-param}). We assumed a generic value of the microturbulent
        velocity    $v_\text{turb}    =    2.0\,\text{km}\,\text{s}^{-1}$.   The
        angle-dependent    emergent   specific   intensities  $I(\lambda,\theta,
        \varepsilon_\text{Si},\varepsilon_\text{Cr},\varepsilon_\text{Fe})$ were
        computed using the SYNSPEC code.

    \subsection{Variability of the radiative flux}

        \begin{table}[t]
            \caption{
            Parameters   of   the Gaussian  functions  used to
                approximate the  transmissivity  of the used filters of the
                ten-colour system.}
            \label{tab-filters}
            \begin{center}
                \begin{tabular}{ccc}
                    \hline
                    Band   &  Central wavelength [\AA]  & Half-width [\AA]  \\
                    \hline
                    $U$    &  3450                     &  200                \\
                    $P$    &  3750                     &  130                \\
                    $X$    &  4050                     &  110                \\
                    $Y$    &  4620                     &  130                \\
                    $Z$    &  5160                     &  105                \\
                    $V$    &  5420                     &  130                \\
                    $H
                    \!
                    R$     &  6000                     &  200                \\
                    $S$    &  6470                     &   50                \\
                    \hline
                \end{tabular}
            \end{center}
        \end{table}

        The intensity $I_c(\theta, \varepsilon_\text{Si}, \varepsilon_\text{Cr},
        \varepsilon_\text{Fe})$ in the  photometric  colour $c$  emerging from
        a stellar    atmosphere    with    abundances
        $\varepsilon_{\text{Si}},
        \varepsilon_{\text{Cr}}$, and $\varepsilon_\text{Fe}$ can be  determined
        by integrating the specific intensity over the wavelengths,
        \begin{equation}
            \label{barint}
            I_c(\theta,\varepsilon_\text{Si},\varepsilon_\text{Cr},
                \varepsilon_\text{Fe})= \int_0^{\infty}\Phi_c(\lambda) \,
            I(\lambda,\theta,\varepsilon_\text{Si},\varepsilon_\text{Cr},
                \varepsilon_\text{Fe})\, \text{d}\lambda,
        \end{equation}
        \noindent  where  $\Phi_c(\lambda)$  is the response function  for 
        colour
        $c$.    We   modelled   the  variability   in   the   colours   \textit{
        U,\,P,\,X,\,Y,\,Z,\,V,\,HR},  and  $S$   of  the  ten-colour photometric
        system~\citep[see ][]{tencolour},  which   we  approximated  by  Gauss
        functions.   The   parameters   of  these  functions  can   be  found in
        Table~\ref{tab-filters}.

        The entire surface of the star was subdivided into a grid of  $68 \times
        33$ elements, which corresponds to the resolution of the abundance maps.
        For each of  these  surface elements,  the corresponding  intensity  was
        calculated  based on  the abundances  taken from  the abundance
        maps  by  means  of a  trilinear interpolation  in the grid of synthetic
        spectra.

        The total  radiative flux  $f_c$ in a colour  $c$  of a star with radius
        $R_*$  detected  by  an  observer  at  the  distance  $D$  at  any given
        rotational   phase   can   be   found   using   the     formula
        \citep{mihalas}
        \begin{equation}
            \label{radflux}
            f_c=\zav{\frac{R_*}{D}}^2\intvidpo I_c(\theta,\Omega)\cos\theta\,
                \text{d}\Omega.
        \end{equation}
        The  magnitude difference in  a colour $c$ is then defined as
        \begin{equation}
            \label{def-mag}
            \Delta m_{c}=-2.5\,\log\,\zav{\frac{{f_c}}{f_c^\mathrm{ref}}},
        \end{equation}
        where ${f_c^\mathrm{ref}}$ is the reference flux determined in
        such  a  way  that,  averaged  over  an  entire  rotational  period, the
        resulting magnitude difference will be equal to zero.

\section{Results}

    \subsection{Influence of chemical elements}

        \begin{figure*}[tp]
            \centering
             \resizebox{0.75\hsize}{!}{\includegraphics{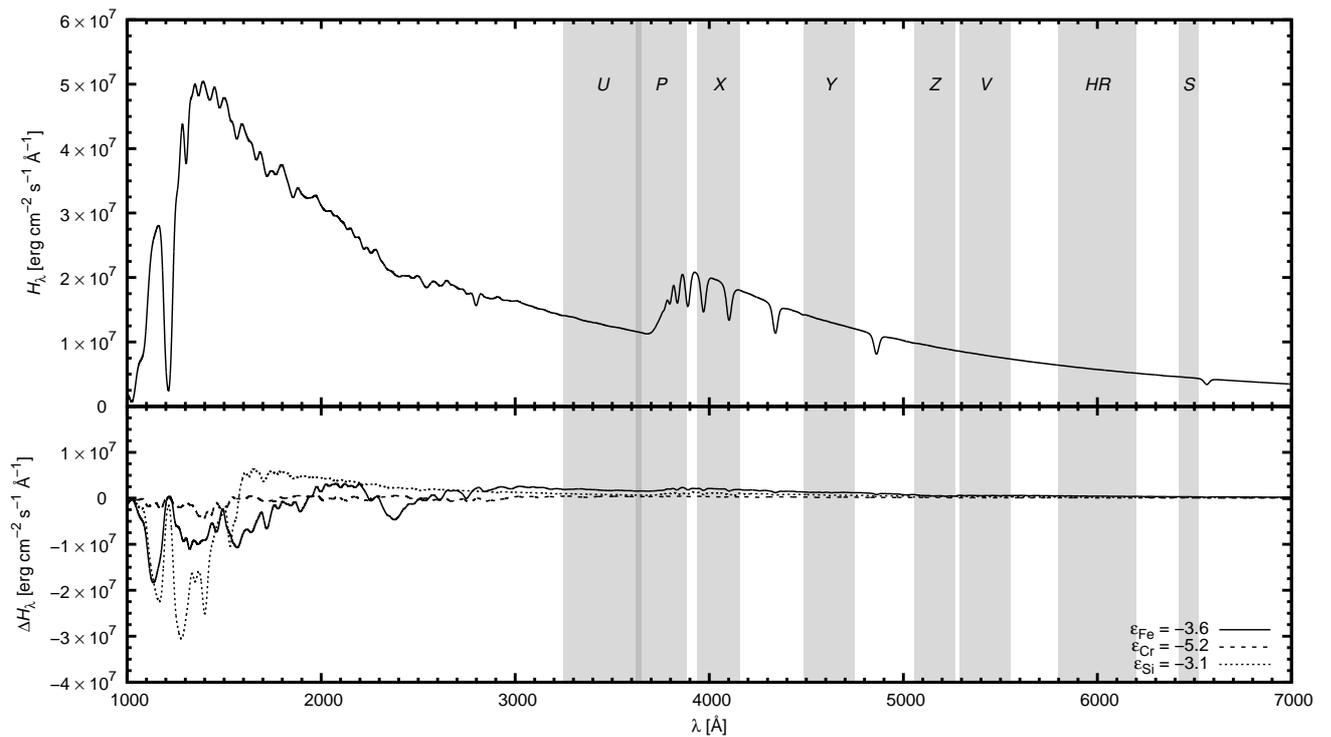}}
            \caption{
                {\em  Upper  plot:}   Emergent   flux  from  a  reference  model
                atmosphere           with          $\varepsilon_\text{Si}=-4.6$,
                $\varepsilon_\text{Cr}=-6.7$, and  $\varepsilon_\text{Fe}=-5.1$.
                {\em  Lower  plot:}  Emergent flux from  the  model  atmospheres
                with  modified abundance of  individual elements  minus the flux
                from the reference model. All fluxes were smoothed by a Gaussian
                filter  with  a dispersion  of $10$\,\AA\ to show the changes in
                continuum,   which  are   important  for  SED  variability.  The
                passbands of the ten-colour photometric  system are shown in the
                graph as grey areas.
            }
            \label{fig-flux_elem}
        \end{figure*}

        \begin{figure}[t]
            \centering
            \resizebox{0.98\hsize}{!}{\includegraphics{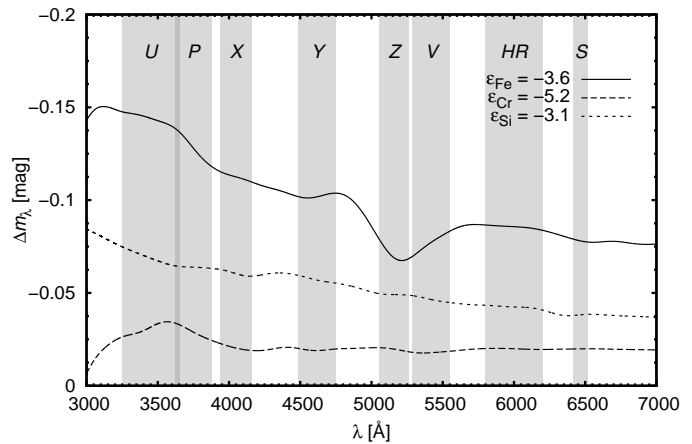}}
            \caption{
                Magnitude  difference  $\Delta  m_\lambda$ between  the emergent
                fluxes calculated  with an enhanced abundance of  the individual
                elements  and  the  reference  flux  $H_\lambda^\text{ref}$ from
                an  atmosphere       with        $\varepsilon_\text{Si} = -4.6$,
                $\varepsilon_\text{Cr}=-6.7$, and  $\varepsilon_\text{Fe}=-5.1$.
                The plots were smoothed with a Gaussian filter with a dispersion
                of $100\,$\AA.
            }
            \label{fig-mag_elem}
        \end{figure}

        The  chemical  elements  present  in  stellar atmospheres  influence the
        way  the  energy  is  distributed  in  the  spectrum.  Bound--bound  and
        bound--free transitions are responsible for the absorption of a
        significant
        portion  of   the  energy  in   some  parts  of  the spectrum.   For the
        sake of the energetic  balance  in the stellar atmosphere,  this  energy
        has  to  be re-emitted  in  other  regions  of  the  spectrum.  This  is
        demonstrated  in  Fig.~\ref{fig-flux_elem},  where  we show the spectral
        energy distribution emergent  from a  reference stellar  atmosphere with
        low abundance of silicon, iron, and chromium and the differences in  the
        energy  distribution  between  the  reference  atmosphere  and   stellar
        atmospheres  with  an  overabundance  of silicon,  iron,  and  chromium.  Higher  abundance  of  heavier  elements leads  to higher
        brightness   in    the    visible    region.    This    is    shown   in
        Fig.~\ref{fig-mag_elem}   where  we  plot   the  magnitude  differences
        between  the  flux  calculated  with  enhanced  abundance of  individual
        elements and the flux emergent from the reference atmosphere.

        Silicon  and  iron  cause  strong  absorption  in the far-UV part of the
        spectrum,  especially between  1100 and 1500\,\AA. The absorption in the
        iron  lines  is  a  characteristic  feature  for  the adjacent UV region
        1500--2000\,\AA. In addition, most of the energy absorbed by silicon  in
        the far-UV  region  1100--1500\,\AA\  is  re-emitted  in the region with
        wavelengths  1500--2000\,\AA.  The  visible  part  of  the  spectrum  is
        mostly  dominated  by  re-emission  of the energy absorbed by  iron
        with  some  contribution  of  silicon.  A  high  concentration  of  iron
        absorption  lines  suppresses  the  contribution  of  iron  to  the light
        variability in a small region  around  5200 \AA. This leads to a typical
        flux depression \citep{kupzenil}.

        Some  influence of chromium is also visible in Figs.~\ref{fig-flux_elem}
        and  \ref{fig-mag_elem},  but  it  is    weaker than the influence of
        silicon  and  iron  as  a  result  of  a relatively low maximum chromium
        abundance.  Consequently,  chromium does not have  a significant impact on
        the general character of the variability of the star.

    \subsection{Variability of the star}

        \begin{figure}[tp]
            \centering \resizebox{0.90\hsize}{!}{\includegraphics{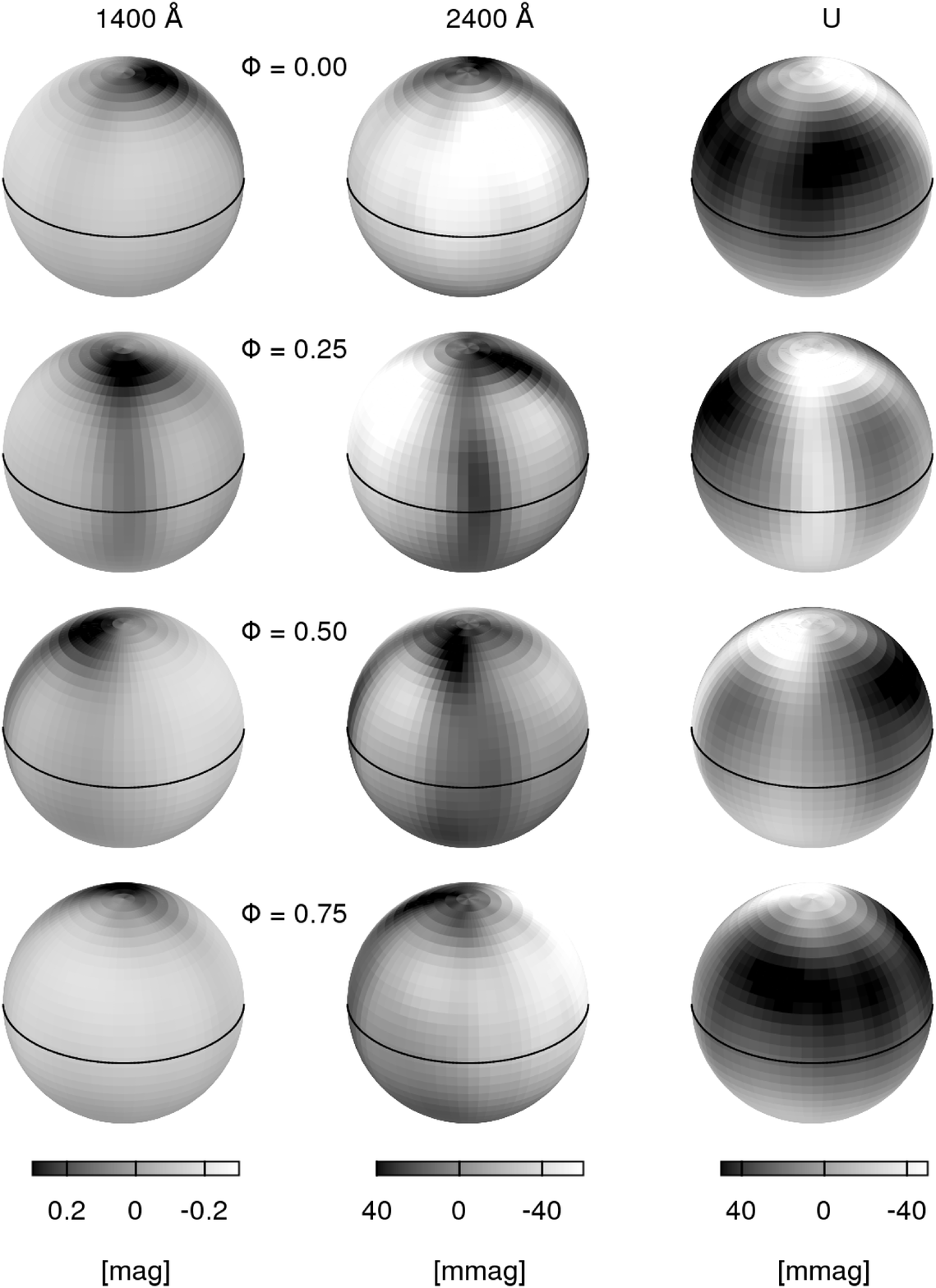}}
            \caption{
                Emergent  intensity  from  the  surface  of  \pdra\  at  various
                rotational  phases  at  wavelengths  of  1400~\AA\  (left panel),
                2500~\AA\ (middle panel), and in the $U$  band  of the ten-colour
                system (right panel). The intensities are displayed as $-2.5\log
                ({I}/{I_0})$,  where  $I_0$  is a  reference intensity chosen in
                such  a  way  that  the  resulting logarithm,  averaged over the
                entire stellar  surface, is equal to zero.}
            \label{fig-fmap}
        \end{figure}

        \begin{figure*}[tp]
            \centering \resizebox{0.90\hsize}{!}{\includegraphics{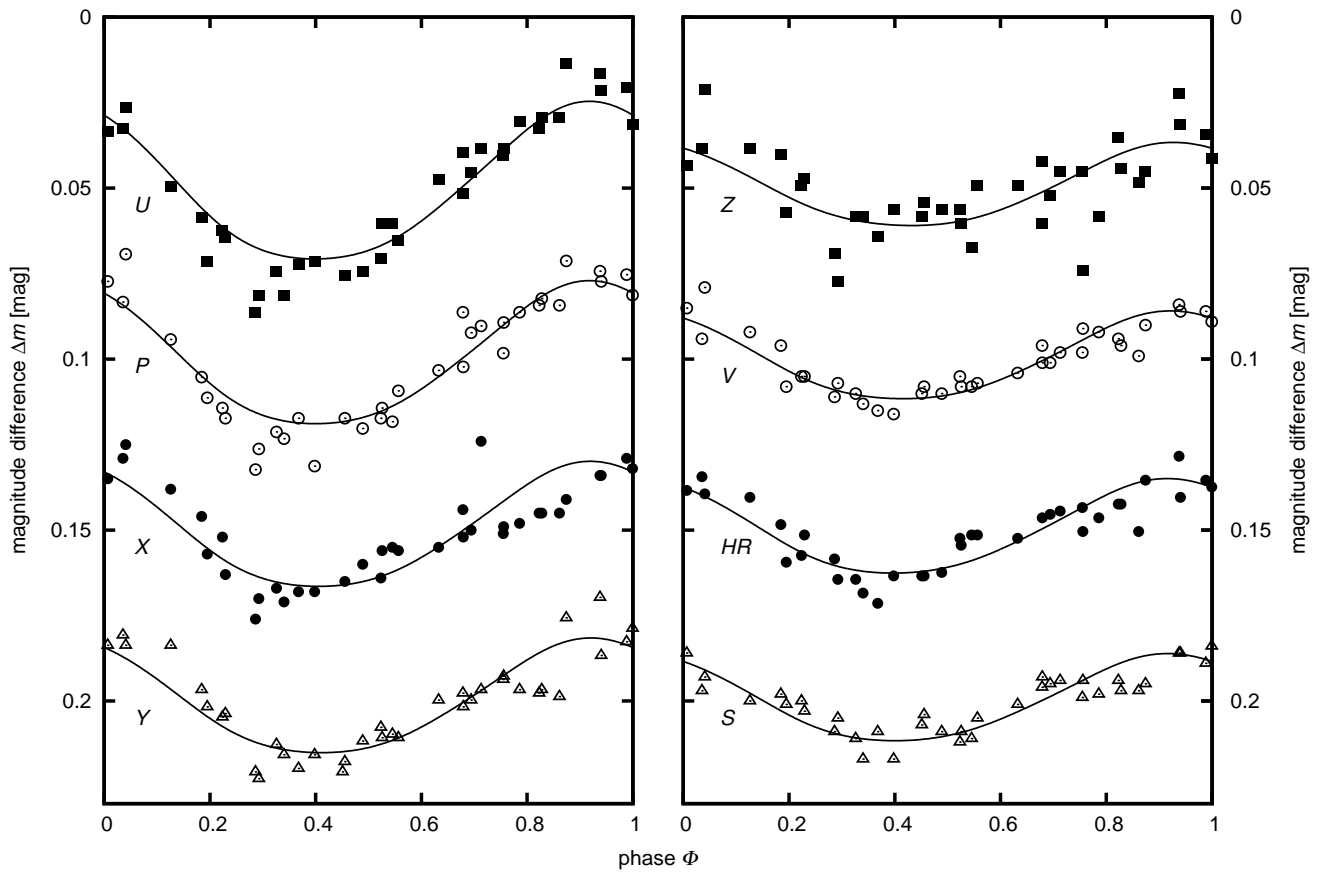}}
            \caption{
                Light   curves   of    \pdra\    in   the   passbands  \textit{%
                U,\,P,\,X,\,Y,\,Z,\,V,\,HR},    and   $S$   of   the  ten-colour
                photometric system. Modelled curves (solid lines)  are  compared
                to  observational  data \citep{musielok1980}.  The  light curves
                have  been  vertically  shifted  to better demonstrate the light
                variability.}
            \label{fig-curves}
        \end{figure*}

        \begin{figure*}[tp]
            \centering
            \resizebox{0.90\hsize}{!}{\includegraphics{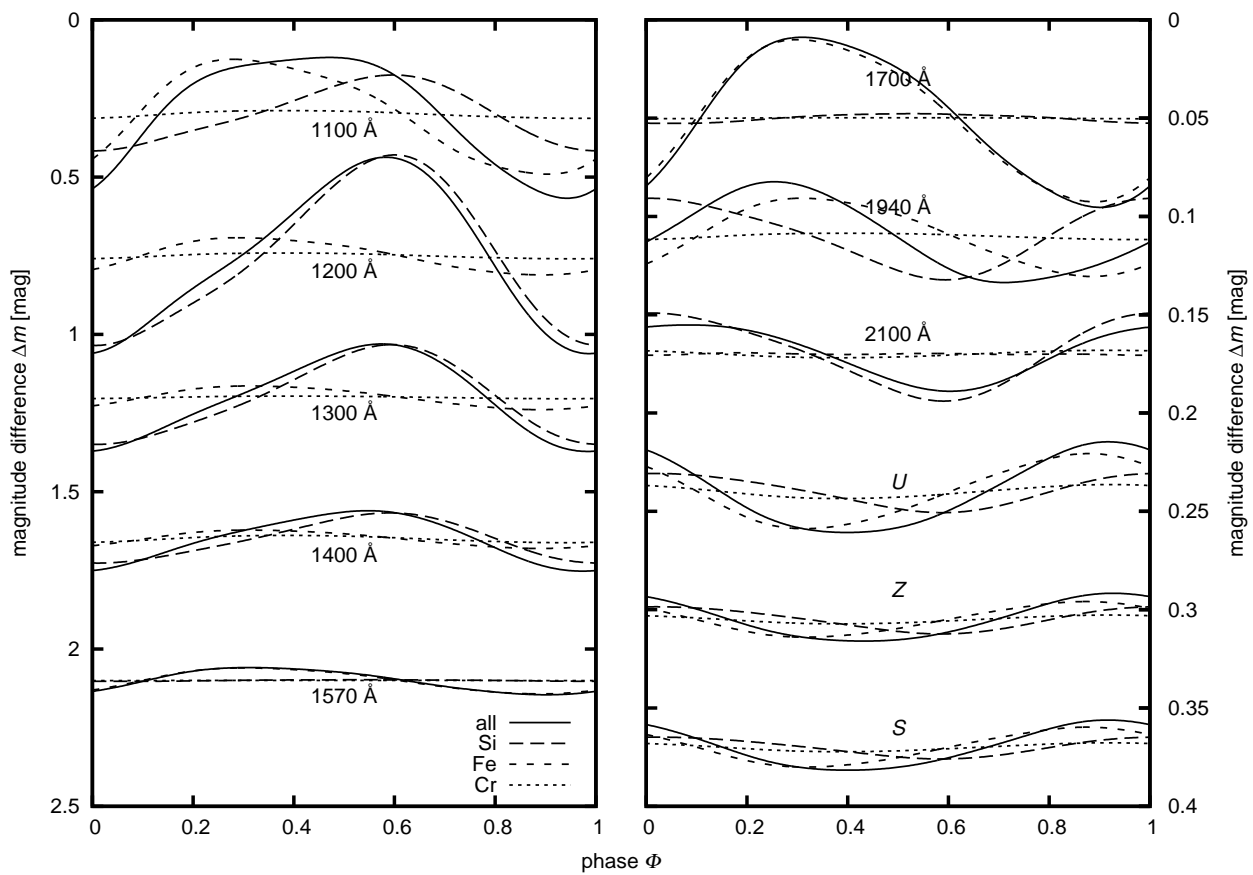}}
            \caption{
                Light curves of \pdra\ synthesised using  the abundance  maps of
                the  individual  elements separately,  compared  to the  overall
                variability  of  the  star  (solid  line)  in  several  Gaussian
                passbands  with  the  dispersion  of $20$~\AA\ in the UV and the
                passbands $U,\,Z$, and $S$ of the ten-colour photometric  system.
                The  light  curves  have  been   shifted  vertically  to  better
                demonstrate the variability. We note  the  very different
                vertical scales of the two panels.}
            \label{fig-components}
        \end{figure*}

        \begin{figure*}[tp]
            \centering \resizebox{0.75\hsize}{!}{\includegraphics{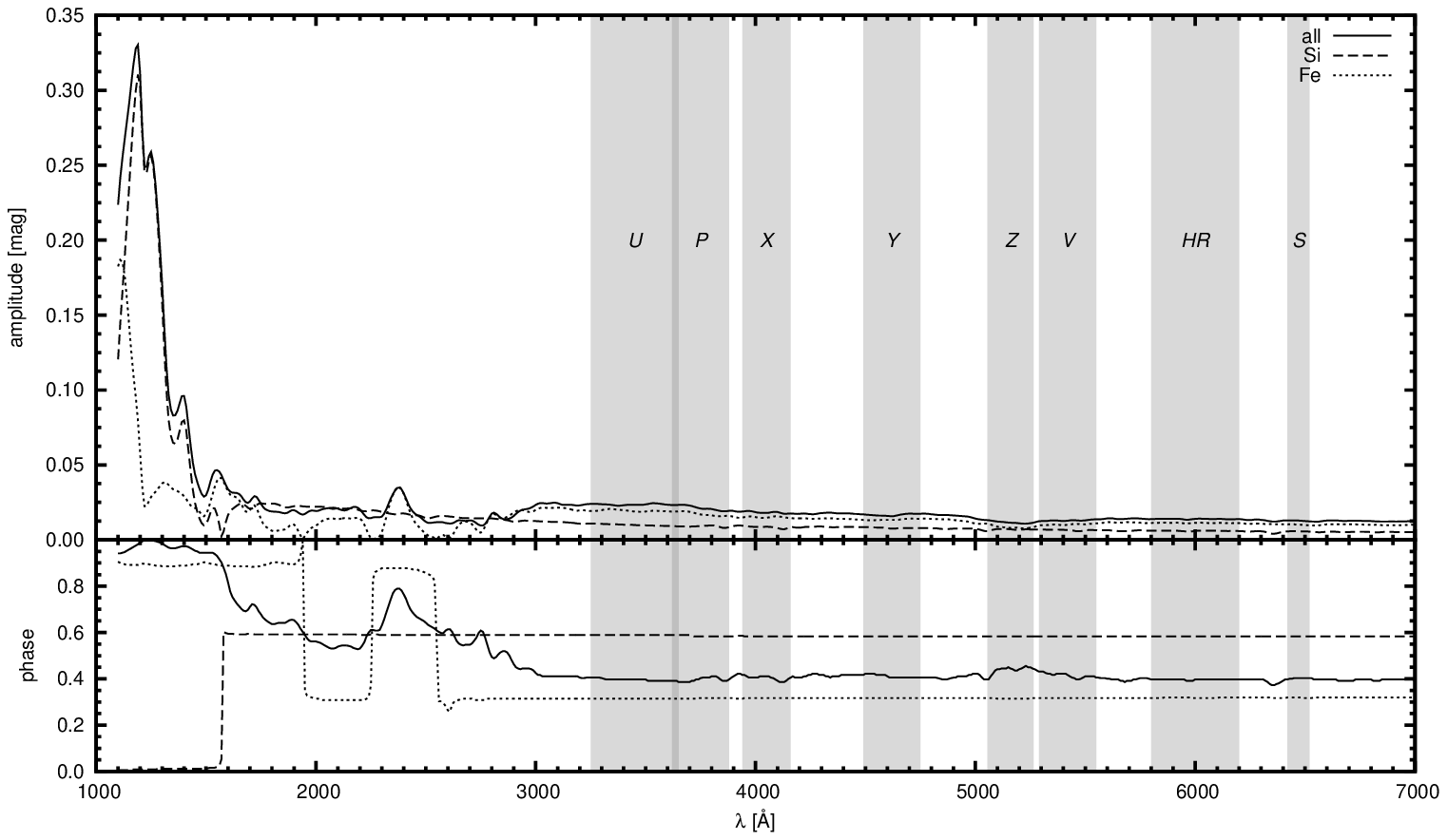}}
            \caption{
                {\em Upper plot:}  Predicted amplitude  (half of the  difference
                between  the maximum  and the minimum  magnitude of the star) of
                the photometric variability  of the star  in Gaussian  passbands
                with  a  dispersion  of  20~\AA\  as   a function of wavelength.
                {\em Lower plot:} The phase of the minimum of the variability of
                the star in the same passbands plotted against the wavelength.}
            \label{fig-amplitudes}
        \end{figure*}

        We model  the  variability  of \pdra{} using abundance maps adopted from
        \citet{kuschnigthesis}.  Both  silicon  and  iron  maps  feature a large
        area  with  a higher abundance  than  the  rest  of  the  surface.
        Figure~\ref{fig-fmap} shows the emergent intensity from the various points
        across the stellar disk at various rotational phases in several spectral
        regions.  Areas with high silicon and iron abundance  appear as dark
        spots
        on the stellar  surface in the far-UV, whereas they appear bright in the
        visible  as  a  result  of  the  flux  redistribution. The opposite holds
        for regions with lower silicon and iron abundance.

        The  light curves  of the star  are shown in Fig.~\ref{fig-curves}.  The
        modelled   curves   are   compared   to  the   photometric  observations
        \citep{musielok1980}. Both the amplitude and the shape of  the  modelled
        curves agree with the observations. The variability of the star
        is most pronounced in the $U$ band, while in the $Z$ band the amplitude
        is the smallest of all modelled curves. The shape of the  curves remains
        roughly the same in all passbands in the  visible  spectrum  and bears a
        strong resemblance to a sine function. The light curve  in the $Z$ band,
        both the modelled and the observed, is slightly shifted in phase
        compared to the other curves.

    \subsection{Effect of the individual elements}

        In  order  to  see  how  the  individual  chemical  elements  affect the
        variability  of  the star,  we synthesised the  light curves of the star
        using only one of  the abundance maps at a time,  keeping the abundances
        of  the  other  elements  constant.  These  light  curves  are  shown in
        Fig.~\ref{fig-components} in selected UV and visual passbands.

        The variability is strongest in the region around  1200\,\AA, mostly due
        to  the  absorption by silicon. The modelled light curve at
        1400\,\AA~qualitatively resembles the observed light curve presented
        by~\citet{jamar1977}, the observed light curve having somewhat larger
        amplitude (approx. 0.15 mag compared to the 0.096 mag amplitude for the
        modelled one). 
        However, since the  original  data  from  the  paper  are  not available
        and important information is missing, such as the ephemeris or the width of the
        passband used,  we cannot sensibly plot the observed light
        curve together with the modelled one. At about 1570\,\AA, we can observe
        the transition between the inversely correlated absorption and
        re-emission regions for silicon. The contribution of silicon to the
        overall variability  of the star at this wavelength is  practically null
        and the light variations are caused almost exclusively by iron
        absorption. An analogous  situation, where the  iron light curve shows
        practically no variations, can be seen at approximately $1940$\,\AA.
        Chromium does not contribute  significantly  to  the  light  curves
        at any wavelength as a result of its relatively low abundance.

        The light curves showing the  variability due to the individual elements
        vary  in  amplitude  and sign,  but  they  retain  their shape and phase
        throughout  the  examined  spectrum.  The  curves are, however,  shifted
        in  phase  relative to each other,  because  the individual elements are
        distributed differently across the stellar surface.

        On the  other hand, the shape of the light curves obtained using all
        abundance maps  changes significantly with the wavelength.  The shape of
        the resulting light curve depends on  the ratio of the amplitudes of the
        components of  the variability  due to the individual  chemical elements
        and their orientation,  i.e. whether the  particular element contributes
        to  the  variability  by  absorption   or  by  re-emission  due  to  the
        backwarming effect.

        While there are some zero-amplitude regions in the  UV  and the visible,
        separating  the absorbing and the re-emitting parts of the spectrum  for
        the  individual  elements, there  is  no  such  region for the resulting
        variability  of the star.  The reason is that the zero-amplitude regions
        for the various examined elements are located  at different wavelengths.
        This  demonstrates  that the concept  of  the null wavelength region,
        i.e. where  no variability shows at a given wavelength,  is an oversimplification
        that may occur only as a result  of  observations  that are not precise
        enough.

        Figure~\ref{fig-amplitudes}  shows  the  predicted  amplitude of the light
        variations in  the narrow-band  passbands  across the UV and the visible
        spectrum.  The amplitude is  greatest  in  the  region  1200--1400\,\AA,
        dropping quickly with  increasing wavelength.
        In   the   lower   plot   of
        Fig.~\ref{fig-amplitudes},  we  show  the  phase position   of the light
        minimum of the variations at various  wavelengths plotted together  with
        the  minima  of the variability  due to silicon and iron.
        The  position of  the minima  of  the variability  caused by  individual
        elements practically do not change,  except for  a few  sudden  changes,
        which actually represent the transitions between  the absorbing and  the
        re-emitting  regions  for  silicon  and  iron,  where  the  curves  flip
        vertically,  so that the minima become maxima and vice versa.  The phase
        of  the  light  minimum  of  the  resulting  light curve drifts fluidly
        between the  minima  of  the  individual  elements,  gravitating  toward
        the one with the greater amplitude.

    \section{Influence of stellar multiplicity}
        
        \begin{figure*}[tp]
            \centering \resizebox{0.75\hsize}{!}{\includegraphics{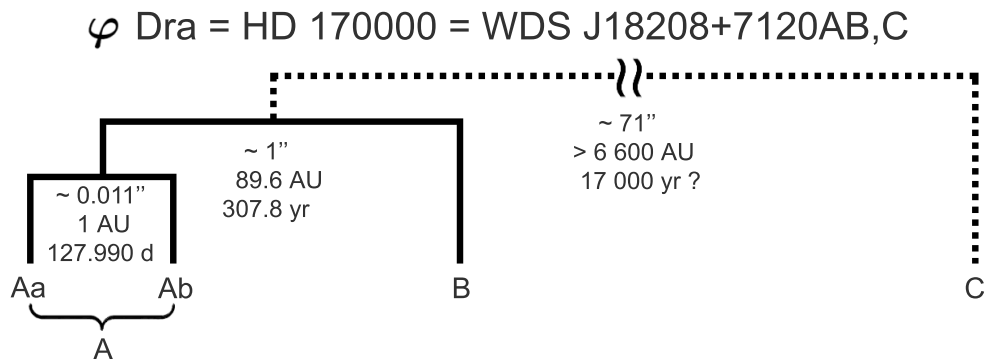}}
            \caption{
                    Diagram of the multiple system \pdra~with expected values
                    for angular and absolute distances between components (total
                    semi-major axes between binary pairs) and their orbital
                    periods~(Li\v ska 2015, in preparation). }
            \label{fig-system}
        \end{figure*}

        The studied object (Aa) is a part of a multiple stellar system \citep[%
        Li\v ska  2015,    in   preparation]{beardsley1969,andrade2005}. It is 
        the brighter component of an SB1-type spectroscopic binary (Aa, Ab) with
        a  more  distant,  visual  companion  (B).  There  is  a possible fourth
        (optical) component C presented in the  Washington Double Star Catalogue
        \citep{mason2001}. A diagram of the system is given in
        Fig.~\ref{fig-system}.
        
        The multiplicity of the system may affect the light curve modelling. We
        therefore investigate the   influence of the other components on our
        results.

    \subsection{Influence of light time effect}

        Each of the components  influences the observed period  of CP
        variability by the light time effect (LiTE), which could be visible in
        the \textit{O-C} diagram. Semi-amplitude $A_{\rm Aa}$ of LiTE can be
        calculated directly from RV changes \citep{beardsley1969}. We obtained
        $A_{\rm Aa} \sim 0.0012$  day $\doteq 104$ sec with orbital period
        (127.990(4) days, Li\v ska 2015, in preparation)  of  system  Aab. This
        is  under the accuracy of available measurements. The B  component
        causes higher LiTE  amplitude $A_{\rm Aab}\sim0.2$ day (maximum value).
        This value is relatively high, but the whole cycle takes several
        hundred  years (e.g. Andrade 2005 estimated an orbital period of 307.8
        years). Consequently,  our  analysed observations in a 40-year interval
        were influenced only slightly,  so the rotational  period  can  be  accurately
        approximated  by  a  constant   value
        $P=1\fd716500(2)$ (see Fig. \ref{OC}).

    \subsection{Influence of RV changes}

        \citet{kuschnigthesis}  used  spectra  for  DI  obtained  during  5 days
        in term JD$_\mathrm{hel}$ 2449797.404--2449802.684  that correspond to
        the orbital phases 0.05--0.09 of the inner system Aab determined with
        elements adopted  from  Li\v ska   2015  (in   preparation).  
        The semi-amplitude of the radial velocity of the  inner orbit of the Aa
        component  was  found to be $K_{\rm Aa} = 29.0(1.3)$\,km\,s$^{-1}$.
        Therefore, the radial velocity of the Aa component would  have  changed
        during  Kuschnig's  observations by almost  11\,km\,s$^{-1}$;  for a wavelength of 5000\,\AA, this would roughly correspond to a  shift  of
        $\Delta\lambda_{5000} = 0.18$\,\AA.  This is  low in  comparison  to the
        rotational   velocity   of   the   star   (95\,km\,s$^{-1}$,   see Table
        \ref{tab-param}).
        Changes in RV due to the visual B component in this short term interval
        is also negligible due to the long period and  low semi-amplitude
        $K_{\rm Aab} \sim 5$\,km\,s$^{-1}$ (maximum value).

    \subsection{Contribution of the components to the flux}

        The  B  star  is  fainter  than  Aab   in the Hipparcos band   by  about
        $\Delta H_\mathrm{p} = 1.445(10)$\,mag   \citep{esa97},   in the   Tycho
        bands $\Delta  B_\mathrm{T} = 1.50(1)$\,mag,
        $\Delta  V_\mathrm{T} = 1.42(1)$\,mag
        \citep{fabricius2000},  and  in red and infrared
        $\Delta R = 1.42(2)$\,mag, $\Delta I  =  1.33(3)$\,mag
        \citep{rutkowski2005}.    Flux   ratios  $F_{
        \rm Aab}/F_{\rm B}$ in these bands  are  0.264  ($H_\mathrm{p}$),  0.251
        ($B_\mathrm{T}$),  0.270  ($V_\mathrm{T}$),  0.270  ($R$),  0.294 ($I$).
        From  these  values,  it is evident that 21\%--23\% of the total flux in
        optical  bands  belongs  to  the  B  component. The observed  amplitude
        of brightness variations for the Aab component is about 21--23\% higher
        after subtracting the B component flux. We do not have enough
        information about the Ab component, but we assume  its  contribution  to
        the total flux is negligible.  The  influence  of  the  B  component  on
        the  photometric variability is not negligible,  but it is probably
        partially compensated by its influence on the line strengths and derived
        abundances. The influence of the B component may be one of the reasons
        why the observed amplitude of the flux variations at 1400 \AA~is slightly
        larger than the predicted amplitude.

\section{Discussion}

    There  are  several  simplifications  that  can  affect  our results. In our
    computations, we neglected the effects of the magnetic  field of  the  star.
    Phi Draconis has  a   moderate   magnetic   field   with  a  poloidal  component
    $B_{\mathrm{p}}    \approx  3$~kG~\citep{landstreet1977}.    However,     as
    demonstrated  by  \cite{khan2006},  magnetic  fields  of  this  magnitude do
    not have significant impact on the SED variability of CP stars.

    The  abundance  maps  were  derived  using  a  mean stellar atmosphere, as
    opposed to computing  local model atmospheres for the individual elemental
    abundances present on  the  stellar surface. This means the influence of the
    modified,  varying  chemical  composition  on the  internal structure of the
    atmosphere  was  not taken  into account during the  mapping.  According  to
    some studies \citep[e.g.][]{stift2012}, this  introduces  an  error  in the
    computations, which  may result  in incorrect abundance maps.  However,  the
    good agreement  between our  models and the observations indicates  that the
    accuracy  of the abundance maps is sufficient. The detailed tests of Doppler
    mapping techniques also support this conclusion \citep{kowas}.

\section{Conclusions}
    
    Using our own data, as well as the archive photometric data, we have found  a new
    linear  ephemeris  for the light maximum based on  Eq.~\eqref{efem}  in the
    optical region of stellar spectra. There are no signs of period inconstancy
    of the star.
    
    We  successfully  modelled  the SED variations of the star \pdra\ in the  UV
    and the  visible  regions. The assumption  of  an  inhomogeneous  horizontal
    surface distribution of heavier  elements,  together  with  spectral  energy
    redistribution and the rotation of the star, can fully  explain the observed
    light variations.

    The variability of the star is  caused  mainly  by  bound--bound transitions
    of  iron  and  bound--free  transitions  of  silicon. While the influence of
    silicon  is  more  significant in the near and far UV spectrum,  the visible
    region is mostly dominated by   re-emission due to  the presence of iron.
    Chromium  also  contributes  to  the  light  variations, but its role in the
    variability of the star is  much less  significant than  that  of  iron  and
    silicon.

    The variability of the star is caused by  multiple  chemical  elements, with
    different  horizontal  distributions,  which  has  several  consequences. The
    general shape of the light  curve  strongly  depends on the  wavelength. The
    light curves  in the various spectral regions are shifted in phase  with respect
    to each other.  There  is  no  zero-amplitude  region  in  the  UV or in the
    visible, separating the absorbing and the re-emitting parts of the  spectrum
    with inversely correlated variability.

    Our  models  nicely  predict  the  variability  of  the star in the visible,
    including both the shape and the amplitude of the light variations. However,
    our  results may  be  slightly influenced by the multiplicity of the system.
    The models also predict the slight shift in phase for the light curve in the
    $Z$ band of the ten-colour system, which agrees with observations.
    The lack of quantitative UV observations  of
    the star makes it is impossible for us to verify the correctness of our models in
    this part of the spectrum. This is very unfortunate  because the  processes
    that are  most  interesting  for the variability of the star occur in this
    particular region.

    This  work  indicates  that  the  variability  of most CP stars is caused by
    inhomogeneous surface distribution  of  heavier elements and rotation of the
    star.  It is a strong argument in favour of the correctness  and accuracy of
    the abundance maps, the atomic data, and the model atmospheres.

\begin{acknowledgements}
    The   authors   of   this  paper  acknowledge  the  support  of  the  grants
    GA \v CR P209/12/0217,
    M\v SMT 7AMB14AT015 (WTZ CZ 09/2014), and in final phases of GA\,\v{C}R
    13-10589S.
    This work  was supported  by the Brno Observatory and Planetarium.
    TL acknowledges support by the Austrian FFG within ASAP11 and by the FWF NFN
    projects S11601-N16 and S116 604-N16.
\end{acknowledgements}

\end{document}